\documentclass[a4paper,UKenglish,cleveref, autoref, thm-restate]{lipics-v2021}
\hideLIPIcs

\usepackage{hyperref}
\usepackage{soul}
\usepackage{graphics}
\usepackage{graphicx}
\usepackage{multicol}
\usepackage{xcolor}
\usepackage[vlined,ruled,linesnumbered,noresetcount]{algorithm2e}
\usepackage{subcaption}
\usepackage{float}

\newcommand{\remove}[1]{}

\newcommand{\TryLock}{\mbox{\textsc{TryLock}}}
\newcommand{\RunAndUnlock}{\mbox{\textsc{RunAndUnlock}}}

\newcommand{\Unlock}{\mbox{\textsc{Unlock}}}
\newcommand{\CreateDescriptor}{\mbox{\textsc{CreateDescr}}}

\newcommand{\Create}{\mbox{\textsc{Create}}}
\newcommand{\Retire}{\mbox{\textsc{Retire}}}

\newcommand{\RetireDescriptor}{\mbox{\textsc{RetireDescr}}}

\newcommand{\Run}{\mbox{\textsc{RunDescr}}}
\newcommand{\CommitValue}{\mbox{\textsc{CommitValue}}}
\newcommand{\CV}{\mbox{\textsc{CommitValue}}}

\newcommand{\FetchUpdatedValue}{\mbox{\textsc{FetchValue}}}
\newcommand{\FV}{\mbox{\textsc{FetchValue}}}

\newcommand{\Recover}{\mbox{\textsc{Recover}}}

\newcommand{\RFlock}{\mbox{\textsc{RFlock}}}

\newcommand{\Flock}{\mbox{\textsc{Flock}}}

\newcommand{\lred}[1]{{\color{red} #1}}
\newcommand{\lblue}[1]{{\color{blue} #1}}
\newcommand{\lblack}[1]{{\color{black} #1}}

\newcommand{\y}[1]{{\color{brown} #1}}

\newcommand{\qedsymb}{\hfill{\rule{2mm}{2mm}}}

\SetKw{Continue}{continue}
\SetKw{Break}{break}
\SetKw{OR}{or}
\SetKw{AND}{and}
\SetKw{NOT}{not}
\SetKwProg{myproc}{Procedure}{}{}

\DontPrintSemicolon

\let\oldnl\nl
\newcommand{\nonl}{\renewcommand{\nl}{\let\nl\oldnl}}

\makeatletter
\newcommand{\removelatexerror}{\let\@latex@error\@gobble}
\makeatother

\newcommand{\recover}{\mbox{\sc Recover}}

\newcommand{\ld}{\mbox{\sc{Load}}}
\newcommand{\store}{\mbox{\sc{Store}}}
\newcommand{\allocate}{\mbox{\sc{Create}}}
\newcommand{\retire}{\mbox{\sc{Retire}}}
\newcommand{\CAM}{\mbox{\sc{CAM}}}

\newcommand{\SaveLogs}{\mbox{\sc SaveLogs}}
\newcommand{\RestoreLogs}{\mbox{\sc RestoreLogs}}

\newcommand{\NULL}{\mbox{\sc Null}}

\newcommand{\CAS}{\mbox{\textit{CAS}}}
\newcommand{\True}{\mbox{\texttt{true}}}
\newcommand{\False}{\mbox{\texttt{false}}}

\newcommand{\pwb}{\mbox{\texttt{pwb}}}
\newcommand{\pfence}{\mbox{\texttt{pfence}}}
\newcommand{\psync}{\mbox{\texttt{psync}}}

\newcommand{\comnospace}{\mbox{$\triangleright$}}
\newcommand{\com}{\mbox{\comnospace\ }}

\renewcommand{\paragraph}[1]{\smallskip\noindent{\bf #1}}

\Copyright{}
\title{Recoverable Lock-Free Locks} 

\author{Hagit Attiya}{Technion -- Israel Institute of Technology, Israel}{ hagit@cs.technion.ac.il}{https://orcid.org/0000-0002-8017-6457}{Supported by the Israel Science Foundation (22/1425 and 25/1849)}

\author{Panagiota Fatourou}{FORTH ICS, Greece and University of Crete, Computer Science Department}{ faturu@csd.uoc.gr}{https://orcid.org/0000-0002-6265-6895}{Supported by the Hellenic Foundation for Research and Innovation (HFRI) under the ``Second Call for HFRI Research Projects to support Faculty Members and Researchers'' (Project: PERSIST, number: 3684), and by the Greek Ministry of Education, Religious
Affairs and Sports call SUB 1.1 -- Research Excellence Partnerships (Project: HARSH, code: Y$\Pi$ 3TA-0560901), implemented through the National Recovery and Resilience Plan Greece 2.0 and funded by the European Union -- NextGenerationEU.}

\author{Eleftherios Kosmas}{Hellenic Mediterranean University, Greece}{ ekosmas@hmu.gr}{https://orcid.org/0000-0002-9552-5664}{Supported by the Hellenic Foundation for Research and Innovation (HFRI) under the ``Second Call for HFRI Research Projects to support Faculty Members and Researchers'' (project number: 3684).}

\author{Yuanhao Wei}{University of British Columbia, Canada}{yuanhaow@cs.ubc.ca}{https://orcid.org/0000-0002-5176-0961}{}

\authorrunning{Attiya, Fatourou, Kosmas and Wei}



\category{} 

\relatedversion{} 

\nolinenumbers 

\begin{document}

\maketitle

\begin{abstract}
This paper presents the first transformation that introduces both lock-freedom and recoverability. 
Our transformation starts with a lock-based implementation, 
and provides a recoverable, lock-free substitution to lock acquire 
and lock release operations. The transformation supports nested locks 
for generality and ensures recoverability without jeopardising the correctness of the lock-based implementation
it is applied on.
\end{abstract}

\section{Introduction}
\label{sec:intro}

Novel \emph{non-volatile memory} (NVM) devices is an emerging memory technology, 
currently co-existing with DRAM in some machines.
NVM persists during transient failures and opens the possibility
of designing data structures and applications that can recover their
original state after such failures and continue their execution.
Despite intensive research,
\emph{efficient recoverability} remains a challenge,
due to the intricate persistence behavior of cached information,
and it requires costly synchronization instructions.
To address this challenge, 
many hand-crafted algorithms were proposed
(e.g.,~\cite{DBLP:conf/opodis/BerryhillGT15, ChenQ-VLDB2015, DBLP:conf/ppopp/FriedmanHMP18, CoburnCAGGJW-Asplos2011, RamalheteCFC19, SP21, FPR19, ZF+19}), 
and even general techniques to turn
existing applications into recoverable ones
(e.g.,~\cite{DBLP:conf/wdag/IzraelevitzMS16, DBLP:journals/corr/abs-1806-04780, FB+20, FPR21, AB+20}).

In the context of multi-processing and concurrent programs,
an even older trend is to strive for \emph{lock-freedom}~\cite{Herlihy1991wait},
in order to tolerate delays and (non-transient) failure of threads.
Lock-freedom has been the gold standard for concurrency,
and is provided by numerous algorithms and implementations.
More importantly, there have been several proposals to automatically
derive lock-free implementations.
The classical idea for making algorithms lock-free,
dating back to~\cite{DBLP:conf/spaa/Barnes93,Turek1992locking},
relies on having a thread that takes a lock on an object,
assign an \emph{operation descriptor} to that object.
This descriptor is then used to help the operation complete
in case the object is needed for another operation.
(See more details in Section~\ref{sec:related}.)

Interestingly, neither of these two types of transformations,
adding recoverability or adding lock-freedom,
achieves both goals.
This is unfortunate since recoverability is particularly attractive
when combined with lock-freedom.
This is also curious since similar techniques are used in many of these
transformations, albeit for different purposes.
For example, carefully \emph{tracking} the execution of operations
in order to support \emph{helping} in the case of lock-free transformations,
or to support re-execution of operations in the case of recoverability.
Another aspect is relying on \emph{idempotent} code that can be run
multiple times but appears to have run once~\cite{DeKruijfSJ2012static},
either due to helping (in lock-free transformations) or
due to recovery (in recoverability).

This paper presents the first direct transformation that introduces both
lock-freedom and recoverability.
Our transformation starts with a lock-based implementation,
and produces recoverable, lock-free substitutions
to \emph{lock acquire} and \emph{lock release} operations.
This includes \emph{nested} calls to acquire locks while holding
other locks.
The recoverable, lock-free substitutions ensure the same properties
in the resulting implementation.

Lock replacement is a particularly attractive avenue for
deriving recoverable lock-free applications,
due to their wide applicability.
Our transformation builds on \Flock~\cite{ben2022lock}, 
a recent highly-optimized implementation of lock-free locks. 
Their main departure from the classical approach
of~\cite{DBLP:conf/spaa/Barnes93,Turek1992locking}
is in the way idempotence is ensured.
Instead of tracking the progress of code through object descriptors,
they use a \emph{read log} shared by all threads helping the same code.
The log tracks all reads from shared mutable locations,
and is modified with CAS.
This ensures that all threads executing the same code use
the same values read,
which in turn, mean they execute the same writes.
To increase the applicability of their transformation,
they support nesting of locks,
by making the locking code itself be idempotent.

Our transformation, called \RFlock, starts with \Flock,
since it has an explicit handling of idempotence
and 
comes with an efficient implementation.
However, several key changes are needed in order to introduce 
recoverability.
The most important one is to maintain \emph{update logs}, 
in addition to read logs.
Before applying any update on shared memory, 
the update is first recorded in the log.
The log is persisted before the updates are applied.

Another important aspect is that \RFlock\ avoids in-place updates,
since this would require to log and persist each update,
at a significant persistence cost~\cite{attiya2021tracking}.
Instead, \RFlock\ defers the persistence
and the application of updates to the end of a critical section.

Summarizing, the contributions of this paper are:
\begin{itemize}
\item We present a general transformation of lock-based implementations to their recoverable lock-free analogs.
\item We extend \Flock\ to get a new approach that provides recoverability in addition to lock-free try-locks. 
\item The proposed approach supports nested locks for generality and wide applicability. 
\item The transformation preserves the correctness of the lock-based data structure and ensures recoverability on top.  
\end{itemize}

\section{Definitions}
\label{sec:pre}

We consider a system of asynchronous crash-prone \emph{threads} which
communicate through  \emph{base objects} supporting atomic read, write,
and Compare\&Swap (CAS) \emph{primitive} operations.

We assume that the main memory is non-volatile, 
whereas the data in the cache or registers are volatile.
Thus, writes are persisted to the non-volatile memory using 
explicit flush instructions, or when a cache line is evicted.
Under \emph{explicit epoch persistence}~\cite{DBLP:conf/wdag/IzraelevitzMS16},
a write-back to persistent storage is triggered by a
\emph{persistent write-back (\pwb)} instruction;
a \pwb{} flushes all fields fitting in a cache line.
The order of \pwb s is not necessarily preserved.
A \pfence{} instruction orders preceding \pwb s
before all subsequent \pwb{}s.
A \psync\ instruction waits until all previous \pwb s
complete the write backs.
For each location, persistent write-backs preserve program order.
We assume the \emph{Total Store Order} (\emph{TSO}) model,
supported by the x86 and SPARC architectures,
where writes become visible in program order.

The concurrent program implements a collection of \emph{methods}.
A thread $\mathit{q}$ \emph{invokes} a method to start its execution;
the method \emph{completes} by returning a \emph{response value},
which is stored to a local variable of $\mathit{q}$
(and thus it is lost if a crash occurs before $\mathit{q}$ persists it).

At any point during the execution of a method,
a system-wide \emph{crash-failure} (or simply a \emph{crash})
resets all volatile variables to their initial values.
Failed threads are recovered by the system asynchronously,
independently of each other;
the system may recover only a subset of these threads
before another crash occurs.

We assume methods synchronize through locks. 
The code section protected by a lock is called a 
\emph{thunk}~\cite{Ingerman1961thunks},
and it is a function that takes no arguments, 
and returns a boolean value, 
indicating whether the thunk was successful.
To synchronize, threads call \TryLock{}, 
with the corresponding Lock and thunk as parameters.
A typical code would look as in Figure 1.

\begin{figure}
\label{fig:critical}
\begin{flushleft}
1: \hspace*{.5cm} \TryLock($L_1$, [=] \{	\\
2: \hspace*{1cm}     SOME CODE \\
3: \hspace*{1cm}     \TryLock($L_2$, [=] \{ \\
4: \hspace*{1.5cm}         SOME CODE \})\\
5: \hspace*{.5cm}\})
\end{flushleft}
\caption{Typical code with \TryLock.}
\end{figure}

In this code, Lines~2--4 are the thunk of the \TryLock{} in Line~1, 
while Line~4 is the thunk of the \TryLock{} in Line~3.
Note that lock acquire and release are implicit in this programming style. 
Also, this style also implies \emph{pure} nesting, i.e., 
the scope of locks is properly nested within each other. 
As in~\cite{ben2022lock}, we borrow the notation [=] for lambdas in C++ to denote a thunk; 
a lambda allows the programmer to define an anonymous function object 
right at the location where it is invoked or passed as an argument to a function. 

An execution is {\em durably linearizable}~\cite{IMS16}, 
if the effects of all methods that have completed before a crash 
are reflected in the object's state upon recovery.
A recoverable implementation is \emph{lock-free},
if in every infinite execution produced by the implementation,
which contains a finite number of system crashes,
an infinite number of methods complete.

\section{Overview and Interface of RFlock}
\label{sec:rflock-over}

\noindent
{\bf Overview.}
Our transformation utilizes \Flock~\cite{ben2022lock},
which transforms a lock-based implementation
into a lock-free one, by enabling threads to help other
threads to execute their critical sections.
In case a thread $p$ wants to acquire a lock that is currently
held by another thread $p'$, then instead of blocking,
$p$ helps $p'$ by executing the entire critical
section of $p'$. Then, $p$ unlocks the lock and can compete for the lock again
in order to continue with the execution of its critical section.

We first describe the key elements of \Flock.
A lock is implemented as a pointer to a descriptor
where all necessary information is stored to allow helping.
To ensure \emph{idempotent} execution of a critical section
(by multiple threads),
\Flock\ maintains a {\em read log} of the values
read during each read access within a critical section.
The first thread to perform such a read records the read value in the log.
Subsequent threads will read these values by accessing the read log
and thus, they will follow the same execution path within the critical section.
Therefore, they will all perform the same updates of memory locations;
the latter are atomically applied in place using \CAS.
A pointer to the log is stored in the descriptor associated
with the lock at each point in time.
A boolean variable is also stored in the lock to indicate
whether the lock is acquired.
Thus, the {\em value} of a lock is a vector containing the
pointer to the lock descriptor and the value of the boolean variable.

Recall that a \emph{thunk} is a function that contains
the code of the critical section,
takes no arguments, and returns a boolean value.
The use of thunks supports helping,
as a helper of a critical section
can simply invoke the thunk.

To support \emph{nesting} of critical sections,
\Flock{} logs also the values of locks.
Consider  a critical section $\mathit{CS'}$,
nested within a critical section $\mathit{CS}$,
and let $L$ be the lock protecting $\mathit{CS'}$.
Every thread that wants to acquire $L$,
it attempts to log $L$'s value in the read log of $\mathit{CS}$.
Only one of them succeeds.
Then, all threads use the logged value to acquire $L$. After acquiring $L$,
a thread continues with the execution of $\mathit{CS'}$ (using the read log
of $\mathit{CS'}$ which is different than that of $\mathit{CS}$).
In case a helper thread attempts to acquire $L$ after
it has already been taken, then its acquisition effort will fail,
but it may continue with the execution of $\mathit{CS'}$
(in case it is not already completed).

\RFlock\ builds upon the main idea of \Flock\
to support the idempotent execution of
critical sections by maintaining read logs.
However, the design of \RFlock\ departs from \Flock\
by avoiding in-place updates.
In-place updates would require to log and persist each update individually,
which would result in significant persistence cost~\cite{attiya2021tracking}.
To enhance its performance, \RFlock\ defers the persistence and the application
of updates to the end of the thunk. 

\RFlock\
maintains an {\em update log} per lock, in addition to its read log to support recoverability.
Before applying any update on shared memory, the update is first recorded in the update log.
The update log is persisted before the updates are applied.

Deferring updates to the end of a critical section, can result in executions where reads following a write
in a critical section may have different responses than in the original code $\mathcal{C}$, on which \RFlock\ is applied.
Whether this will happen  depends on the memory consistency model of the programming language $\mathcal{L}$,
in which $\mathcal{C}$ has been written, and the level of consistency chosen by the programmer.
In cases where the memory consistency model of $\mathcal{L}$
provides consistency guarantees as strong as sequential consistency, then \RFlock\ requires
the following assumption to ensure correctness.
{\em For each thunk, no read on a shared location occurs after the first write into such a location.}
This assumption is needed to avoid the following bad scenaro.
Consider two shared variables A and B and assume that the critical section of a thread
$p_1$ first writes A and then reads B, whereas the critical section of a thread $p_2$
first writes B and then reads A. If the consistency memory model of $\mathcal{L}$
ensures sequential consistency, one of the two reads will return the new written value in $\mathcal{C}$.
This is the case, for instance, when the writes to A and B are protected by locks and the
memory consistency model ensures that acquiring or releasing a lock causes all writes
to be flushed.
Since \RFlock\ defers updates, in the absence of the assumption above,
in \RFlock, these
reads could return the old values for both A and B.

\RFlock\ is correct even when the assumption 
does not hold
in cases where the same lock protects the two writes to A and B in $\mathcal{C}$.
Moreover, if no locks protect these two writes in $\mathcal{C}$, then the scenario described above
can appear even when running $\mathcal{C}$, thus the behavior of \RFlock\ is then correct.
In cases where the two writes are protected by different locks,
the memory consistency model of many programming languages, including C++,
allows this scenario to occur even in executions of the original code $\mathcal{C}$. Thus, \RFlock\ is also correct then.
Summarizing, whether the assumption 
is needed for proving \RFlock\ correct depends on
the level of consistency the programmer chooses for $\mathcal{C}$ from the levels supported by the programming language
(through its memory consistency model), and has to be checked separately in each case.
\RFlock\ can also support in-place updates at a higher persistence cost; the {\em persistence cost}
is the cost that \RFlock\ pays (at executions with no crashes) to execute
the necessary persistence instructions in order to provide recoverability in case of crashes.
Then, \RFlock\ would be correct without requiring that the assumption 
holds.

\RFlock\ does not support I/O within a critical section. In order for lock-free locks~\cite{ben2022lock} to ensure lock-freedom,
the code inside a critical section needs to complete within a finite number of steps, so condition variables
(or any form of waiting) cannot be used inside critical sections.
The same holds for \RFlock.

\noindent
{\bf Interface.}
\RFlock\ uses the same application interface as \Flock. We illustrate this interface
with an example of a simple lock-based implementation of a concurent queue~\cite{MichaelS-PODC1996},
whose code is provided in Algorithm~\ref{alg:concurrent-queue}.
The implementation uses two locks, a lock called TailLock, to synchronize the enqueuers,
and another called HeadLock, to sychronize the dequeuers. The queue is implemented by a linked
list whose first element is a dummy node. Variables Head and Tail are shared, as well as the fields
of the nodes of the list that implements the queue. Only next fields of nodes are updated,
whereas the rest of the fields are immutable.

Algorithm~\ref{alg:flock-example} shows how this code is written using the interface of \Flock.
The calls to Lock() on lines~\ref{alg:cq-lock-enq} and~\ref{alg:cq-lock-deq} are substituted by TryLocks (lines~\ref{alg:rflock-lock-enq} and~\ref{alg:rflock-lock-deq}), which call the
appropriate thunks. In the thunks, whenever variables Head and Tail are read or written,
this is done by calling \ld\ and \store, respectively.
The same is true with the next fields of the nodes of the linked list.
Allocation of memory should happen in the thunk (line~\ref{alg:rflock-alloc}).

In \Flock, the assignment operator has been overloaded, so that calling \store\ in the application code in not needed.
The same can be done in \RFlock\ to simplify the task of incorporating the appropriate interface in the application.

Our transformation works on top of any lock-based implementation.
A programmer needs to simply rewrite those parts of its code
that use locks utilizing the \RFlock\ interface described here. After ensuring that the original code has been rewritten
using the \RFlock\ interface appropriately, the obtained implementation  provides similar correctness guarantees as the original lock-based implementation,
in addition to being durable linearizable and lock-free.
Debugging needs to be performed solely on the original lock-based code.

\begin{algorithm}[t]
	\nonl
	\SetKwBlock{Begin}{}{}	
	\removelatexerror
	\footnotesize
	\begin{flushleft}
		type Node \{ \hspace*{13.25mm}  \;
		\hspace*{6mm} Int $\mathit{key}$ \;
		\hspace*{6mm} Node *$\mathit{next}$ \;
		\}\;
		Lock HeadLock, TailLock \hspace*{2cm}{\em initially, released}\;
		Node *Head, *Tail \hspace*{2.9cm}{\em initially, pointing to a dummy node}\;
	\end{flushleft}


\nl	\myproc{void {\footnotesize Enqueue}(Int $\mathit{key}$)}{
\nl		Node *nd := new<Node>($\mathit{key}$, NULL) \;
\nl		Lock(TailLock) \label{alg:cq-lock-enq}\;
\nl		Tail $\rightarrow$ next := nd\;
\nl		Tail := nd\;
\nl		Unlock(TailLock)\;
}
\nl	\myproc{Int {\footnotesize Dequeue}(void)}{
\nl		Int result \;
\nl		Lock(HeadLock) \label{alg:cq-lock-deq}\;
\nl		\uIf{Head $\rightarrow$ next = \NULL}{
\nl			result := EMPTYQUEUE	
		}
\nl		\uElse{
\nl			result := Head $\rightarrow$ next $\rightarrow$ $\mathit{key}$\;
\nl     Node *oldHead = Head\;
\nl			Head := Head $\rightarrow$ next\;
\nl     Retire(oldHead);
		}
\nl		Unlock(HeadLock)\;
\nl		return result\;
	}
\caption{Simple Lock-based Concurent  Queue Implementation from~\cite{MichaelS-PODC1996}.}
\label{alg:concurrent-queue}
\end{algorithm}

\begin{algorithm}[t]
	\nonl
	\SetKwBlock{Begin}{}{}	
	\removelatexerror
	\footnotesize
	\begin{flushleft}
		type Node \{ \hspace*{13.25mm}  \;
		\hspace*{6mm} Int $\mathit{key}$ \;
		\hspace*{6mm} mutable Node * $\mathit{next}$ \;
		\}\;
		Lock HeadLock, TailLock \hspace*{2cm}{\em initially, released}\;
		mutable Node *Head, *Tail \hspace*{1.7cm}{\em initially, pointing to a dummy node}\;
	\end{flushleft}


\nl	\myproc{void {\footnotesize Enqueue}(Int $\mathit{key}$)}{
\nl		\TryLock(TailLock, [=] \{ \label{alg:rflock-lock-enq} \;
\nl	\hspace*{.5cm}	Node *nd := \allocate<Node>($\mathit{key}$, NULL) \label{alg:rflock-alloc}\;
\nl	\hspace*{.5cm}		Node *tail := Tail.\ld()\;
\nl	\hspace*{.5cm}		tail $\rightarrow$ next.\store(nd)\;
\nl	\hspace*{.5cm}		Tail.\store(nd)\;
\nl	\hspace*{.5cm}		return \True\;
		\}) \;
}

\nl	\myproc{Int {\footnotesize Dequeue}(void)}{
\nl		Int result \;
\nl		\TryLock(HeadLock, [=] \{ \label{alg:rflock-lock-deq}\;
\nl	\hspace*{.5cm}		Node *head := Head.\ld()\;
\nl	\hspace*{.5cm}		Node *headNext := (head $\rightarrow$ next).\ld()\;
\nl	\hspace*{.5cm}		\uIf{headNext = \NULL}{
\nl		\hspace*{.5cm}			result := EMPTYQUEUE
	\hspace*{.5cm}		}
\nl	\hspace*{.5cm}		\uElse{
\nl	\hspace*{.5cm}			result := headNext $\rightarrow$ $\mathit{key}$\;
\nl	\hspace*{.5cm}			Head.\store(headNext)\;
\nl	\hspace*{.5cm}			Retire<Node>(head)\;
	\hspace*{.5cm}		}
\nl	\hspace*{.5cm}		return \True\;
		\})	\;
\nl		return result\;
	}
\caption{Example of using \Flock\ on the implementation of Algorithm~\ref{alg:concurrent-queue}. }
\label{alg:flock-example}
\end{algorithm}

\begin{algorithm}[tb]
	\nonl
	\SetKwBlock{Begin}{}{}	
	\removelatexerror
	\footnotesize
	\begin{flushleft}
		type Log is shared<$\mathit{entry}$>$\mathit{[logSize]}$ \;
		type Thunk is function with no arguments returning bool \;
		\vspace*{1mm}
		\com Private local variables of process $q$: \;
		Log* $\mathit{log_q}$[READ,UPDATE\lred{,LOCK}], initially all $\NULL$ \tcp*{current logs}
			\tcp*{\lblue{$\mathit{log_q}$[READ] maintains $\mathit{\langle V \rangle}$ entries}}
			\tcp*{\lblue{$\mathit{log_q}$[UPDATE] maintains $\mathit{\langle shared\langle V \rangle}$*$\mathit{, V, V \rangle}$ entries}}
			\tcp*{\lred{$\mathit{log_q}$[LOCK] maintains $\mathit{\langle Lock}$*$\mathit{, descriptor}$*$\rangle$ entries}}						
		int $\mathit{pos_q}$[READ,UPDATE\lred{,LOCK}], initially all $0$ \tcp*{current positions in logs}
		\lblue{Set<V*> *$\mathit{retSet_q}$, initially pointing to an empty set \;}
		\vspace*{1mm}

		type mutable<V> \{ \;
 \hspace*{.5cm}		shared<V> $\mathit{val}$ \;
 \hspace*{.5cm} \myproc{V {\footnotesize \ld}()}{	
 \hspace*{.5cm}		V $\mathit{v}$ := \FetchUpdatedValue($\mathit{\&val}$) \label{alg:rec-no-nest:mutable:fetch}\;
	\hspace*{.5cm}	$\mathit{\langle retVal , - \rangle}$ := \CommitValue($\mathit{v}$, READ) \label{alg:rec-no-nest:mutable:commit}\;
	\hspace*{.5cm}	\KwRet $\mathit{retVal}$ \label{alg:rec-no-nest:mutable:load:return}\;
				}
 \hspace*{.5cm} \myproc{void {\footnotesize \store}(V $\mathit{newV}$)}{	
 \hspace*{.5cm}		V $\mathit{oldV}$ := \FetchUpdatedValue($\mathit{\&val}$) \label{alg:rec-no-nest:mutable:store:load}\;
 \hspace*{.5cm}	    \CommitValue($\mathit{\langle \&val, oldV, newV \rangle}$, UPDATE) \label{alg:rec-no-nest:mutable:store:commit}\;
				}
 \hspace*{.5cm} \myproc{void {\footnotesize \CAM}(V $\mathit{oldV}$, V $\mathit{newV}$)}{	
 \hspace*{.5cm}		V $\mathit{check}$ := this.\ld() \label{alg:rec-no-nest:mutable:CAM:load}\;
 \hspace*{.5cm}		\lIf {$\mathit{check} \neq \mathit{oldV}$}{\KwRet} \label{alg:rec-no-nest:mutable:CAM:if}
 \hspace*{.5cm}		$\mathit{val}.\CAS(oldV, newV)$		 \label{alg:rec-no-nest:mutable:CAM:cas} 
	}
\} \ \ \tcp{end of type mutable<V>}
	\end{flushleft}


	\myproc{$\mathit{V}$ {\footnotesize \FetchUpdatedValue}(V *$\mathit{pv}$)}{
\nl		\uIf {$\mathit{log_q}$[UPDATE]$\neq$ \NULL\ \AND\ $\mathit{pv} \in \mathit{log_q}$[UPDATE]}{
\nl			int $\mathit{lpos} =$ last entry containing $\mathit{pv} \in \mathit{log_q}$[UPDATE]\;
\nl			\uIf  {$\mathit{lpos} \leq \mathit{pos_q}$[UPDATE]}{
\nl				\KwRet $\mathit{newVal}$ of $\mathit{log_q}$[UPDATE][$\mathit{lpos}$] \; 
			}
		}
\nl		\KwRet *$\mathit{pv}$
	}


	\myproc{$\mathit{\langle V, boolean \rangle}$ {\footnotesize \CommitValue}(V $\mathit{val}$, \{READ,UPDATE\lred{,LOCK}\} Type)}{
\nl		V $\mathit{cval}$\;
\nl		\lIf {$\mathit{Type}$ = READ}{ $\mathit{cval} := val$} 
\nl		\lElse{$\mathit{cval} := \bot$}
\nl		\lIf {$\mathit{log_q[Type]} = \NULL$}{ \KwRet $\mathit{\langle cval, \True \rangle}$}		
\nl		int $\mathit{pos}$ = $\mathit{pos_q}$[Type] \label{alg:commit:pos} \;
\nl		boolean $\mathit{isFirst}$ := $\mathit{log_q[Type][pos].\CAS(\bot, val)}$ \label{alg:primitives-no-nest:commit:cas}\;
\nl		V $\mathit{returnVal}$ := $\mathit{log_q[Type][pos]}$ \label{alg:primitives-no-nest:commit:read}\;
\nl		$\mathit{pos_q[Type]}$ := $\mathit{pos}+1$\;
\nl		\KwRet $\mathit{\langle returnVal, isFirst \rangle}$ \label{alg:primitives-no-nest:commit:return}\;
	}

\nl	\myproc{V* {\footnotesize \allocate}<V>($\mathit{args}$)}{
\nl		V* $\mathit{newV}$ = AllocateInNVM<V>($\mathit{args}$)\; 
\nl		$\mathit{\langle obj, isFirst \rangle}$ := \CommitValue($\mathit{newV}$, READ)\;
\nl		\lIf {\NOT\ $\mathit{isFirst}$}{sysFree<V>($\mathit{newV}$)}
\nl		\KwRet $\mathit{obj}$\;
	}

	\myproc{{\footnotesize \retire}<V>(V* $\mathit{obj}$)}{
\nl		$\mathit{\langle - , isFirst \rangle}$ := \CommitValue($1$, READ)\;
\nl			\lIf{$\mathit{isFirst}$} {\lblue{add $\mathit{obj}$ to *$\mathit{retSet_q}$}}
	}
\caption{Mutable object. 
Code for thread $\mathit{q}$.}
	\label{alg:rec-no-nest:mutable}
\end{algorithm}

\section{Detailed Description of RFlock}

\subsection{Logs and Mutable Objects}
\label{sec:rflock-dt}
Algorithm~\ref{alg:rec-no-nest:mutable} presents different types of objects utilized by \RFlock.
Code in red is required to support nesting and is discussed in Section~\ref{sec:nesting}.

Each thread $q$ maintains three {\em private} variables, namely $\mathit{log_q}$[READ],
$\mathit{log_q}$[UPDATE], $\mathit{log_q}$[LOCK], which store pointers to the read, update and lock logs
of the lock that $q$ is interested in at the current point in time
(i.e., it either competes for it or it has acquired it).
We will call them  {\em read, update and lock log pointers} of $q$.
When $q$ executes a thunk to help another thread finishing a critical section with lock $L$,
it will run the code of $L$'s thunk from scratch using $L$'s logs. To do this, $q$'s 
log pointers will be set to point to $L$'s logs. Then, $q$ will traverse these logs starting from their first position,
using its $pos_q$ variables.
If any of the $\mathit{log_q}$ pointers is \NULL\ at some point in time, $q$ is not
executing a critical section.
Type {\em Log} indicates an array of $\mathit{logSize}$ elements
implementing a log.
In $\mathit{log_q}$[LOCK], $q$ records the locks that acquires during the execution of a thunk
and is needed to support nesting.

Each shared memory location that is accessed
within a thunk (and it is protected by a lock) should be an object of type \emph{mutable}
(Algorithm~\ref{alg:rec-no-nest:mutable}).
A mutable object $O$ stores a value ($\mathit{val}$) of type $V$ and supports the functions: a) \ld() that returns $O$'s current value,
b) \store($\mathit{newV}$) that writes the value $\mathit{newV}$ into $O$, and
c) \CAM($\mathit{oldV},\mathit{newV}$) that atomically compare $O$'s current value with $\mathit{oldV}$ and if they are the same modifies $O$'s value
to $\mathit{newV}$.
For generality, in \RFlock, a mutable object may also be accessed by a thread $q$ outside a critical section. Then,
the $\mathit{log_q}$ pointers are \NULL.

Briefly, \ld\ first discovers the current value of $O$ by calling function \FV\
and then commits the value read in the appropriate log by calling function \CV;
\store\ behaves similarly.
\FV\ takes as an argument a pointer to the $val$ field of $O$ and returns the current value of $O$.
Its argument is needed to identify whether $O$ has been written
in previous steps of the executed thunk, which means that its value is recorded in the corresponding update log
and should be taken from there, or whether it should be taken directly from the $\mathit{val}$ field of the object.
\CV\ takes as an argument the value $\mathit{val}$ to commit in the appropriate log and a boolean $\mathit{Type}$
that identifies in which of the three logs the value should be committed. It returns the value
that is actually recorded for the operation that called \CV\ in the log. It also returns a boolean
value which is used only for the purpose of properly deallocating objects. Allocation and deallocation
of objects is performed using \Create\ and \Retire.
We next provide the details of the implementation. 

\noindent{\bf LOAD.}
Consider a thunk $th$ protected by a lock $L$
and assume that a thread $q$ calls an instance $ld$ of \ld\ for the $k$th time during the execution of $th$.
Suppose that $q$ has performed $m$ accesses to shared memory (including allocations of shared variables) by the invocation of $ld$.
Assume that \ld\ is invoked to read the value of object $O$.
\ld\ reads the value of $O$ by calling \FV, and attempts to record this value in the $k$th position of the read log of $L$
by calling \CommitValue.

\FetchUpdatedValue\ copes with the following cases.
Either $q$ has already invoked \store\ for $O$ (lines 52-54),
in which case the value of $O$ is the value parameter of the last \store\ performed by $q$ on $O$ thus far,
or it is the curent value recorded into $O$ (line 55).

After \FV, $q$ calls \CV\ attempting to record the value it gets as a parameter in the $k$th position of the log (line 60). 
Its attempt may fail since helper threads may have recorded a value in this position already.
In this case, $q$ returns the value that is already recorded there  (lines 61 and 63). 

\RFlock\ copes also with the case that an instance of \ld\ or \store\ is called
outside a critical section (for generality reasons).
\CommitValue\ addresses such calls to \ld\ and \store\ with lines 56-58. 
Specifically, \CV\ discovers that its pointer to the corresponding log is NULL in this case
and returns either $\bot$ in case of a \store\ 
or the current value of $O$ in case of a \ld. 

Recall that $q$ runs the code of $th$ from scratch using its log pointers (which point to $L$'s logs)
and its local variables $pos_q$, as reflected in the codes of \FV\ and \CV.

We next describe the code of \CV\ in more detail.
When a thread $q$ invokes \CommitValue() with parameters $\mathit{val}$ and $\mathit{Type}$,
$q$ attempts to record the value $\mathit{val}$ into
the next available position of the log pointed to by $\mathit{log_q}[\mathit{Type}]$, using a \CAS\ (line~\ref{alg:primitives-no-nest:commit:cas}).
This position is identified by $\mathit{pos_q[Type]}$ (line~\ref{alg:commit:pos}).
Other threads may concurrently try to record values at the same position of the log.
\CommitValue\ returns  the value of the winner thread (lines~\ref{alg:primitives-no-nest:commit:read}, \ref{alg:primitives-no-nest:commit:return}),
i.e., the value written by the single successful \CAS\ on this position of the log.

\noindent{\bf STORE.}
A $\mathit{store}$ logs a pointer to $O$ in the update log,
together with the value returned for $O$ by  \ld\ (line~\ref{alg:rec-no-nest:mutable:store:load}) and the new value
to be stored in it.
It does so by calling \CommitValue\ (line~\ref{alg:rec-no-nest:mutable:store:commit}) with parameter $\mathit{Type}$
equal to UPDATE.

\noindent{\bf CAM.}
\CAM\ is only needed to appropriately implement the acquisition of a nested lock.
Recall that the lock value may need to be recorded in the read log (to support nested locks),
so the lock is implemented as a mutable object.
Different threads may try to acquire the lock at the same time by storing into it
the descriptor they have created locally. \CAM\ provides the necessary functionality for ensuring
that only one of them succeeds. The rest of the threads use the descriptor recorded in the log.
Specifically, a $\mathit{CAM(old,new)}$ atomically updates the value of $O$ from
$\mathit{old}$ to $\mathit{new}$ using a \CAS\ instruction (line~\ref{alg:rec-no-nest:mutable:CAM:cas}).
To enhance its performance, 
the value of $O$ is loaded and compared against $\mathit{old}$
(lines~\ref{alg:rec-no-nest:mutable:CAM:load}-\ref{alg:rec-no-nest:mutable:CAM:if})
and only if they are the same, the \CAS\ is executed.
If the \CAS\ of line~\ref{alg:rec-no-nest:mutable:CAM:cas} of \CAM\ is executed
successfully, we say that the \CAM\ is {\em successful}.

\noindent{\bf CREATE.}
A thread that calls \allocate\ allocates a new object of type V in NVM  (line 64). 
It then calls \CV\ to attempt to record the pointer to the new object in the read log pointed to by its read
log pointer. Note that it might be that many threads may attempt to also allocate the same object
and record a pointer to its local copies in the read log. Only one of them will succeed,
which takes back the value \True\ for $\mathit{isFirst}$.
The rest see \False\ in $\mathit{isFirst}$, use the logged pointer as the pointer to the new object (lines 65, 67)
and deallocate the object they allocated themselves (line 66). To avoid leaks after a crash,
a persistent allocator, such as ralloc~\cite{WH+20}, can be used.

\noindent{\bf RETIRE.}
\retire\ attempts to record a value other than $\bot$ (in our case $1$) in the read log of lock $L$.
Only one of the threads will succeed. \CV\ will return the value \True\ for $\mathit{isFirst}$ to this thread.
This thread undertakes the task to actually
declare as retired the object $obj$ on which \Retire\ is invoked.
Idempotence ensures that all other threads that call \Retire\ for $obj$ attempt to write
in the same position of this log (and thus they fail).
The actual garbage collection needs to be performed by the original code
(i.e., \Retire\ should be appropriately invoked in the thunks and a garbage
collection algorithm should be employed by the original application for memory reclamation).
The garbage collector presented in~\cite{WH+20} allows to trace and recycle
unreachable blocks upon recovery from a crash.

\RFlock\ supports the use of randomization in critical sections, but
the random bits need to be logged and persisted
in a way similar to how \Create\ handles memory allocation.

Note that \allocate, \retire, \FV\ and \CV\ are routines that are not necessarily called on a mutable object.
On the contrary, \ld, \store\ and \CAM\ are part of the interface of a mutable object.

\subsection{Descriptor Objects and Critical Sections}
\label{sec:decr}

For each critical section, there is a {\em descriptor} object (Algorithm~\ref{alg:rec-no-nest:primitives}),
which stores pointers to the critical section's 
logs,
as well as to its thunk.
A descriptor object also contains
a boolean variable $\mathit{done}$, initialized to \False, which indicates whether this critical section
has been executed;  $\mathit{done}$ is updated to \True\ at the end of the critical section's execution.
Algorithms~\ref{alg:rec-no-nest:primitives} and~\ref{alg:rec-no-nest:logs} provide pseudocode for the descriptor object. Code in red
copes with nested locks and is discussed in Section~\ref{sec:nesting}.


\begin{algorithm}
	\nonl
	\SetKwBlock{Begin}{}{}	
		
	\removelatexerror
	\footnotesize
	
	\begin{flushleft}
		type descriptor \{ \;
		\hspace*{6mm}	Log* $\mathit{log}$[READ,UPDATE\lred{,LOCK}] \;
		\hspace*{6mm}	Thunk* $\mathit{thunk}$ \;
		\hspace*{6mm}	mutable<boolean> $\mathit{done}$ \;
		\lred{\hspace*{6mm} int $\mathit{owner}$ \;}
		\lred{\hspace*{6mm} descriptor* $\mathit{topdescr}$ \;}
		\lred{\hspace*{6mm} Lock* $\mathit{toplock}$ \;}		
		\}				

		Log* $\mathit{RD}[0..N-1]$, initially $[\NULL,\ldots,\NULL]$ \tcp*{stored in NVM}
		\lred{int $\mathit{wth}$, initially $q$ \tcp*{currently executing thread}}
		\lred{descriptor $\mathit{topD}[0..N-1]$, initially $[\NULL,\ldots,\NULL]$\;}
		\lred{Lock $\mathit{topL}[0..N-1]$, initially $[\NULL,\ldots,\NULL]$}
	\end{flushleft}
	

\nl	\myproc{descriptor* {\footnotesize \CreateDescriptor}(Thunk $f$\lred{, Lock *$\mathit{lock}$})} {			
	\begin{flushleft}
\nl		descriptor* $\mathit{descr}$\; 
\nl		Log* $\mathit{log}$[READ, UPDATE\lred{,LOCK}] \;				\label{alg:line:createdescr-start}
	\end{flushleft}
\nl		\lred{\uIf{$\mathit{wth} = q$ \AND\ $\mathit{topD[q]} = NULL$}{
\nl			\lblack{$\mathit{log}$[READ] := \allocate Volatile<Log>()}\;
\nl			\lblack{$\mathit{log}$[UPDATE] := \allocate<Log>()\;}
\nl			$\mathit{log}$[LOCK] := \allocate<Log>()
		}
\nl		\uElse{
\nl			$\mathit{log}$[READ,UPDATE,LOCK] := $\mathit{topD[wth]} \rightarrow \mathit{log}$[READ,UPDATE,LOCK]
		}
		}
\nl		$\mathit{descr}$ := \allocate<descriptor>$(\mathit{log}, f, \False,\lred{\mathit{wth}, \mathit{topD[wth]}, \mathit{topL[wth]}})$\;
\nl		\lred{\uIf{$\mathit{wth} = q$ \AND\ $\mathit{topD[q]} = NULL$}{
\nl			$\mathit{topD[q]} := descr$ \;
\nl			$\mathit{topL[q]} := lock$
		}}
\nl		\KwRet\ $\mathit{descr}$\;					\label{alg:line:createdescr-end}
}


	\myproc{\lblue{boolean} {\footnotesize \Run}(descriptor* $\mathit{descr}$)}{
\nl		boolean $\mathit{returnVal}$\;
\nl \lblack{Log* $\mathit{prevlog}$[READ,UPDATE\lred{,LOCK}] \;
\nl	int $\mathit{prevpos}$[READ,UPDATE\lred{,LOCK}] \;}	
\nl \lblue{Set<V*> *$\mathit{prevretSet}$}\;
	\lred{
\nl		int $\mathit{prevwth} := \bot$\;
\nl		\uIf{$\mathit{descr \rightarrow owner} \neq \mathit{wth}$}{ \label{alg:rec-inner:run:log:if}		
\nl			$\mathit{prevwth}$ := $\mathit{wth}$\;
\nl			$\mathit{wth}$ := $\mathit{descr \rightarrow owner}$\;
\nl			\lblack{\SaveLogs($\mathit{prevlog,prevpos\lblue{,prevretSet}}$)				\label{alg:primitives-no-nest:save-log}\;
\nl			$\mathit{log_q}$[READ,UPDATE\lred{,LOCK}] := $\mathit{descr \rightarrow log}$[READ,UPDATE\lred{,LOCK}]\; \label{alg:primitives-no-nest:run:log-init} 
\nl			$\mathit{pos_q}$[READ,UPDATE\lred{,LOCK}] := $\{ 0, 0\lred{,0} \}$  \label{alg:primitives-no-nest:run:updatePos-init}\;}
\nl			$\mathit{descr := descr \rightarrow topdescr}$\;
		}
	}
\vspace*{.01cm}
\nl		$\mathit{returnVal}$ := $\mathit{descr \rightarrow thunk()}$ \label{alg:primitives-no-nest:run:thunk}\tcp*{run thunk}
\nl		\lred{\uIf{$\mathit{descr} \rightarrow \mathit{topdescr} = \NULL$}{
\nl			\lblack{\uIf{$\mathit{returnVal} = \True$}{
\nl				\pwb(contents of $\mathit{log_q}$[UPDATE]); \label{alg:rflock-no-nest:rununlock:pwb-updatelog} \hspace*{.5cm}
				\pfence()\;			
\nl				$\mathit{RD[\lblue{wth}] := log_q}$[UPDATE] \label{alg:rflock-no-nest:rununlock:RD}\;
\nl				\pwb($\mathit{\&RD[\lblue{wth}]}$); \label{alg:rflock-no-nest:rununlock:pwb-RD} \hspace*{.1cm}	\psync()\;
\nl				\ForEach {$\mathit{\langle pv, oldV, newV \rangle}$ in $\mathit{log_q}$[UPDATE]} { \label{alg:rflock-no-nest:rununlock:for}
\nl					
					\lIf{$\mathit{\lblue{descr} \rightarrow done.\ld()} = \True$}{\Break}
\nl					$\mathit{pv \rightarrow \CAS(oldV,newV)}$\;
\nl 				$\mathit{\pwb(pv)}$ \label{alg:rflock-no-nest:rununlock:pwb-for}
				}
\nl				\psync()
\vspace*{.01cm}
			}
\nl			$\mathit{descr}$ $\rightarrow \mathit{done.\store(\True)}$ \label{alg:rflock-no-nest:rununlock:done}\;
		
			\lblue{
			\ForEach {$\mathit{obj}$ in $\mathit{retSet_q}$} {
\nl				sysRetire($\mathit{obj}$) \;
				remove $\mathit{obj}$ from *$\mathit{retSet_q}$
			}
			}
		}
\y{			\uIf{$\mathit{returnVal} = \True$}{			
\nl				$\mathit{RD[\lblue{wth}] := \NULL}$ \label{alg:rflock-no-nest:rununlock:RD-reset}; \hspace*{.3cm}
				\pwb($\mathit{\&RD[\lblue{wth}]}$); \hspace*{.3cm}
				\pfence()\; \label{alg:rflock-no-nest:rununlock:log-null} 
			}
	}
	}
	}

	\lred{
		\uIf{$\mathit{prevwth} \neq \bot$}{ \label{alg:rec-inner:run:log:if2}		
\nl			\lblack{\RestoreLogs($\mathit{prevlog,prevpos}$)}					\label{alg:rflock-no-nest:rununlock:restore}\;
\nl			$\mathit{wth}$ := $\mathit{prevwth}$
		}
	}
\nl		\KwRet $\mathit{returnVal}$ \label{alg:primitives-no-nest:run:return}\;
	}
	\caption{Descriptor object. Code for thread $\mathit{q}$.}
	\label{alg:rec-no-nest:primitives}
\end{algorithm}

\begin{algorithm}
	\nonl
	\SetKwBlock{Begin}{}{}	
	\removelatexerror
	\footnotesize

	\myproc{{\footnotesize \RetireDescriptor}(descriptor* $\mathit{descr}$)}{
\nl		\retire<Log>($\mathit{descr \rightarrow log}$[READ])\;
\nl		\retire<Log>($\mathit{descr \rightarrow log}$[UPDATE])\;
\nl		\lred{\retire<Log>($\mathit{descr \rightarrow log}$[LOCK])}\;
\nl		\retire<descriptor>($\mathit{descr}$)\;
	}

	\myproc{{\footnotesize \SaveLogs}(Log *$\mathit{prevlog[]}$, Log *$\mathit{prevpos[]}$\lblue{, Set<V*> *$\mathit{prevretSet}$})}{
\nl		$\mathit{prevlog}$[READ, UPDATE\lred{,LOCK}] $:= \mathit{log_q}$[READ, UPDATE\lred{,LOCK}]	\label{alg:savelogs:1}	\; 
\nl		$\mathit{prevpos}$[READ, UPDATE\lred{,LOCK}] $:= \mathit{pos_q}$[READ, UPDATE\lred{,LOCK}] \label{alg:savelogs:2} \;
\nl		\lblue{*$\mathit{prevretSet} := \mathit{retSet_q}$}	\label{alg:savelogs:3}
	}
	\myproc{{\footnotesize \RestoreLogs}(Log *$\mathit{prevlog[]}$, Log *$\mathit{prevpos[]}$\lblue{, Set<V*> *$\mathit{prevretSet}$})}{
\nl		$\mathit{log_q}$[READ, UPDATE\lred{,LOCK}] $:= \mathit{prevlog}$[READ, UPDATE\lred{,LOCK}] \label{alg:restorelogs:1}\;
\nl		$\mathit{pos_q}$[READ, UPDATE\lred{,LOCK}] $:= \mathit{prevpos}$[READ, UPDATE\lred{,LOCK}] \label{alg:restorelogs:2}\;
\nl		\lblue{$\mathit{retSet_q}$ := *$\mathit{prevretSet}$} \label{alg:restorelogs:3}
	}

	\caption{Descriptor object. Code for thread $\mathit{q}$.}
	\label{alg:rec-no-nest:logs}
\end{algorithm}


\noindent{\bf CreateDescriptor.}
The creator of a desciptor object (lines~\ref{alg:line:createdescr-start}-\ref{alg:line:createdescr-end})
calls \allocate\ to create the 
logs of the object.
Then, it calls \allocate\ one more time to create a new descriptor object,
which stores pointers to the allocated logs
and to the thunk. 
\allocate\ ensures that the created values will be saved in the read log pointed to by $q$'s read log pointer.
This is necessary
for supporting nested locks (and specifically, idempotence in assigning the same descriptor to a nested lock).

\noindent{\bf RetireDescriptor.}
\RetireDescriptor\ calls \retire<V> for the logs and the descriptor.

\noindent{\bf Run.}
When \Run\ is invoked by a thread $q$ with parameter some descriptor $\mathit{descr}$, it executes the thunk (line~\ref{alg:primitives-no-nest:run:thunk}) of the descriptor.
Before doing so, $q$
initializes its local pointers $\mathit{log_q}$[READ], $\mathit{log_q}$[UPDATE] and $\mathit{log_q}$[LOCK]
to point to the logs of the descriptor (lines~\ref{alg:primitives-no-nest:run:log-init}-\ref{alg:primitives-no-nest:run:updatePos-init}).
While executing a thunk, \Run\ accesses mutable memory locations by invoking the \ld\ and
\store\ operations that have been incorporated into the code of the thunk.
The execution of a thunk may end up to be unsuccessful due to contention, in which case it has to be repeated\footnote{
This is, for instance, the case, in an optimistic implementation of a linked list~\cite{DBLP:books/daglib/0020056}[Chapter 9],
where a thread repeatedly
executes attempts to insert or delete an element until it succeeds; the execution of an attempt corresponds to a singe execution of the thunk.
Another example of a doubly-linked list is provided in~\cite{ben2022lock}.
In our simpler example (Algorithm~\ref{alg:flock-example}) all thunks complete successfully.}.
Recall that \RFlock\ executes all updates after the execution of the thunk.
Before doing so, it has a) to persist the contents of the update log of the descriptor (line~\ref{alg:rflock-no-nest:rununlock:pwb-updatelog}),
and b) to persist a pointer to the update log by storing its value in a shared variable $\mathit{RD[q]}$ (line~\ref{alg:rflock-no-nest:rununlock:RD}) and
by persisting $\mathit{RD[q]}$  (line~\ref{alg:rflock-no-nest:rununlock:pwb-RD}). These actions are needed to ensure durable linearizability.
Then, \Run\ continues to apply the updates on shared memory and persist them, one by one
(lines~\ref{alg:rflock-no-nest:rununlock:for}-\ref{alg:rflock-no-nest:rununlock:pwb-for})\footnote{\RFlock{} assumes that the original lock-based code is ABA-free. To make a program ABA-free
a counter can be attached to any shared variable that may suffer from the ABA problem.
The counter should be incremented every time the value of the variable is updated. }.
Next, it sets the $\mathit{done}$ field of the descriptor to \True\
to indicate that the execution of its thunk has been completed (line~\ref{alg:rflock-no-nest:rununlock:done}). 
Finally, $q$ resets its local pointers $\mathit{log}_q$[READ], $\mathit{log}_q$[UPDATE], and $\mathit{log}_q$[LOCK]
(line~\ref{alg:rflock-no-nest:rununlock:restore}),
as well as resets and persists the value stored in $RD[q]$
(line~\ref{alg:rflock-no-nest:rununlock:RD-reset}).

Recall that to help another thread finishing a critical section with lock $L$,
$q$ sets its 
log pointers to point to $L$'s logs.
To be able to continue its own critical section after helping,
$q$ invokes \SaveLogs\ before doing so to get a backup of its log pointers (line~\ref{alg:primitives-no-nest:save-log}).
It also calls \RestoreLogs\ to recover these values as soon as it has finished helping (line~\ref{alg:rflock-no-nest:rununlock:restore}).

\subsection{Flat Locks in \RFlock}
\label{sec:rflock-main}

A lock is implemented as a mutable {\em LockDescr} object
(Algorithm~\ref{alg:rec-no-nest:try-lock}), which contains a pointer to the descriptor of the critical section it protects
and a boolean variable ($\mathit{isLocked}$) that indicates whether the lock is acquired or not.

\begin{algorithm}

	\nonl
	\SetKwBlock{Begin}{}{}	

	\removelatexerror
	\footnotesize
	
	\begin{flushleft}
		type LockDescr \{ \hspace*{13.25mm}  \;
		\hspace*{6mm} descriptor *$descr$ \;
		\hspace*{6mm} Boolean $\mathit{isLocked}$ \;
		\}\;
		type Lock is {\bf mutable}$\mathit{\langle LockDescr \rangle}$\;

	\end{flushleft}


\nl	\myproc{boolean {\footnotesize \TryLock}(Lock *$\mathit{lock}$, Thunk *$\mathit{thunk}$)	}{
\nl		LockDescr $\mathit{CLDescr}$ := $\mathit{lock \rightarrow \ld()}$\;	\label{alg:rflock-no-nest:trylock:load-lock}		
\nl		\uIf{$\mathit{\NOT\ CLDescr.isLocked}$ \lred{\OR\ $\mathit{CLDescr.descr \rightarrow owner} = \mathit{wth}$}} {			\label{alg:rflock-no-nest:trylock:if-unlocked}
\nl			descriptor *$\mathit{descr}$ := \CreateDescriptor($\mathit{thunk}$\lred{$, \mathit{lock}$}) \label{alg:rflock-no-nest:trylock:new-desc}\;
\nl			LockDescr $\mathit{NLDescr}$ := $\langle \mathit{descr}, \True \rangle$\;
\nl			$\mathit{lock \rightarrow CAM(CLDescr, NLDescr)}$ \label{alg:rflock-no-nest:trylock:lock-acq}\;
\nl			$\mathit{CLDescr}$ := $\mathit{lock \rightarrow \ld()}$	\label{alg:rflock-no-nest:trylock:load}\;
\nl			\uIf{$\mathit{NLDescr.descr \rightarrow done}$.\ld() $= \True$ \OR $\mathit{CLDescr} = \mathit{NLDescr}$}{ \label{alg:rflock-no-nest:trylock:if-locked}
\nl				Boolean $\mathit{result}$ := \Run($\mathit{NLDescr.descr}$)\;
\nl				\Unlock($\mathit{lock, NLDescr}$) \label{alg:rflock-no-nest:trylock:run-my-cs} \;
\nl				\lblue{\RetireDescriptor($\mathit{descr}$)\;}
\nl				\KwRet $\mathit{result}$ \label{alg:rflock-no-nest:trylock:return-thunk} \;
			}
\nl			\RetireDescriptor($\mathit{descr}$)
		}
\nl		\Run($\mathit{CLDescr}$)\;  
\nl		\Unlock($\mathit{lock, CLDescr}$) \label{alg:rflock-no-nest:trylock:locked} \;

\nl		\KwRet $\mathit{\False}$	\label{alg:rflock-no-nest:trylock:return} 
	}

\nl	\myproc{boolean {\footnotesize \Unlock}(Lock *$\mathit{lock}$, LockDescr $\mathit{lockDescr}$)}{
\nl	\lred{ 
	descriptor *$\mathit{descr := lockDescr.descr \rightarrow topdescr	}$ \;
\nl	\lIf{$\mathit{descr = \NULL}$}{$\mathit{descr := lockDescr.descr}$}
\nl \uIf{$\mathit{descr \rightarrow done = \True}$}{
\nl			\lblack{LockDescr $\mathit{NLDescr}$}\;
\nl			\ForEach {$\mathit{\langle l, d \rangle}$ in $\mathit{descr \rightarrow log}$[LOCK]} {
\nl				LockDescr $\mathit{LDescr}$ := $\langle \mathit{d}, \True \rangle$\;
\nl				$\mathit{NLDescr}$ := $\langle \mathit{d}, \False \rangle$\;
\nl				$\mathit{l \rightarrow CAM(LDescr, NLDescr)}$
			}
\nl			\uIf{$\mathit{lockDescr.descr \rightarrow topdescr = \NULL}$}{ \lblack{
\nl			$\mathit{NLDescr}$ := $\langle \mathit{lockDescr.descr}, \False \rangle$\;
\nl				$\mathit{lock \rightarrow CAM(lockDescr, NLDescr)}$ \label{alg:rflock-no-nest:rununlock:unlock} \;
				}
				\uIf{$\mathit{descr \rightarrow owner = q}$} {
\nl					$\mathit{topD[q] := \NULL}$ \;
\nl					$\mathit{topL[q] := \NULL}$ \;
				}
			}
			\Else{
\nl				LockDescr $\mathit{LDescr}$ := $\langle \mathit{descr}, \True \rangle$\;
\nl				$\mathit{NLDescr}$ := $\langle \mathit{descr}, \False \rangle$\;
\nl				$\mathit{descr \rightarrow toplock \rightarrow CAM(LDescr, NLDescr)}$\;
			}
	}
	\lElse{
		\CommitValue($\langle \mathit{lock, lockDescr.descr} \rangle$, LOCK)
	}}
	}

\remove{	
	\myproc{{\footnotesize \Unlock}(Lock *$\mathit{lock}$)}{
\nl		LockDescr $\mathit{curLDescr}$ := $\mathit{lock \rightarrow}$ load$()$\;
\nl		LockDescr $\mathit{newLDescr}$ := new LockDescr($\mathit{curLDescr.ldescr}$, $\False$)\;
\nl		$\mathit{lock \rightarrow CAM(curLDescr, newLDescr)}$\;
	}
}
	\myproc{{\footnotesize \recover}()}{
\nl		\ForEach {$\mathit{p \in \{0,1, \ldots, N-1\}}$}{
\nl			Log *$\mathit{updatesLog}$ := $\mathit{RD[p]}$\;
\nl			\uIf{$\mathit{updatesLog} \neq \NULL$}{
\nl				\ForEach {$\mathit{\langle val, oldV, newV \rangle}$ in $\mathit{updatesLog}$} {
\nl					$\mathit{val.\CAS(oldV,newV)}$\;
\nl					$\mathit{\pwb(\&val)}$
				}
\nl				\psync()
			}	
		}
		
\nl		$\mathit{RD[\lblue{q}]}$ := $\NULL$ \label{alg:rflock-no-nest:recover:RD}\;
	}
	\caption{Locks in \RFlock. Code for thread $q$.}
	\label{alg:rec-no-nest:try-lock}
\end{algorithm}

Recall that while executing a critical section, a thread $q$'s log pointers point to
the 
logs, respectively, of the Lock of the critical section;
moreover, $q$'s private variables $\mathit{pos_q}$[]
store the next available position in the 
corresponding logs of the critical section,
where a  new entry should be added. These variables are stored in volatile memory.

\noindent{\bf TryLock.}
Each thread that wants to execute a critical section $\mathit{CS}$, calls the \TryLock\
function 
with two parameters:
a pointer $\mathit{lock}$ to the corresponding Lock and a pointer $\mathit{thunk}$ to the thunk.
\TryLock\ attempts to acquire $\mathit{lock}$ and then it executes the $\mathit{thunk}$.
It returns $\True$ if the $\mathit{lock}$ is acquired and the $\mathit{CS}$ is executed; otherwise,
it returns $\False$. 

In more detail, when a thread $q$ invokes \TryLock, first loads the current value of $\mathit{lock}$ (line~\ref{alg:rflock-no-nest:trylock:load-lock}).
In case no thread currently owns $\mathit{lock}$ (line~\ref{alg:rflock-no-nest:trylock:if-unlocked}), $q$ initializes a new descriptor
for $\mathit{CS}$ (line~\ref{alg:rflock-no-nest:trylock:new-desc}) and then it exectues a CAM, using this descriptor,
to attempt to acquire the $\mathit{lock}$
(line~\ref{alg:rflock-no-nest:trylock:lock-acq}).
The \CAM, in addition to serve the purpose of synchronizing between threads that attempt to execute CS concurrently (as it contains a \CAS),
it is also necessary for supporting nested locks. 
In case $\mathit{lock}$ has been acquired (second condition of line~\ref{alg:rflock-no-nest:trylock:if-locked}),
the $\mathit{thunk}$ is executed by calling \Run.
Then, the $\mathit{lock}$ is unlocked
by calling \Unlock\ (line~\ref{alg:rflock-no-nest:trylock:run-my-cs})
and the value returned by the $\mathit{thunk}$ is returned (line~\ref{alg:rflock-no-nest:trylock:return-thunk}).
The same happens in case the execution of $\mathit{CS}$ has been completed
and its $\mathit{lock}$ has been released by some helper thread (first condition of line~\ref{alg:rflock-no-nest:trylock:if-locked}).

In case some other thread owns $\mathit{lock}$, either because the if statement of line~\ref{alg:rflock-no-nest:trylock:locked}
was evaluated to \False, or because the conditions of the if statement of line~\ref{alg:rflock-no-nest:trylock:if-locked} are evaluated to \False,
then \TryLock\ continues by helping
the thread that has acquired the $\mathit{lock}$ to complete its critical section. This is done by calling \Run\ and \Unlock\ with this thread's descriptor $\mathit{CLDescr}$.
In this case, $\False$ is returned (line~\ref{alg:rflock-no-nest:trylock:return}).

\noindent{\bf Unlock.}
\Unlock\ (lines~170-185) 
unlocks the lock by attempting  to update the Lock's descriptor using \CAM\ (line~\ref{alg:rflock-no-nest:rununlock:unlock}),
so that it contains the value \False\ in the isLocked field.

\noindent{\bf Recover.}
At recovery time, each thread $q$ calls \Recover.
In \Recover, $q$ has to apply all the updates recorded in the update logs of all threads.
This is necessary to ensure durable linearizability in situations, where e.g., we have
a (partially applied) critical section CS for which a crash occurs after at least one
of its updates has been applied and persisted, and before some other update has been persisted.
If $q$ does not ensure that all the updates of CS are applied and persisted,
it may end up to access, in the subsequent thunks it will execute, some of the memory locations that have already been persisted and
others that have not (with obsolete values), thus possibly violating durable linearizability.

In the full version, we argue that 
those updates that have not yet been persisted by the time of the failure, will be applied and persisted 
before the end of (any instance of) \Recover.
Recall that for each thread $p$, $\mathit{RD[p]}$ stores a pointer to the update log of the critical section
that $p$ was executing at the time of the failure.
Because updates are not in-place in \RFlock,
a crash that occurs before any of the updates have taken place
does not jeopardise durable linearizability. 
Moreover, a crash that occurs after $RD[p]$ has been persisted will also not jeopardise
durable linearizability, as all updates performed by $p$ during the execution of its last thunk
are persisted before persisting $RD[p]$.

\subsection{Sketch of Proof}
\label{sec:proof-sketch}

We briefly discuss correctness for \RFlock. Many of our arguments (and our definitions) follow that in~\cite{ben2022lock}.
We first argue that \RFlock\ ensures idempotent execution of thunks.
Consider any execution $\alpha$ and let \textit{th} be any instance of a thunk executed in $\alpha$.
Due to helping, the code of the thunk may have been executed by many different threads.
We argue that all these threads follow the same execution flow while executing \textit{th},
thus they all execute the same code lines of \textit{th}. We also argue that all the threads
read the same value when they read a lock-protected variable and attempt to store the same value when they write such a variable.
For each store, only one of the threads succeeds to record the new value in the update log.
Roughly speaking, if we start from any configuration $C$ at which no invocation of the thunk is active,
we let any number of inactive threads invoke the thunk starting from $C$, and consider all possible executions in which at least one thread completes
the execution of the thunk, at the first configuration at which the execution of the
thunk completes in all of these executions, the values of the shared variables protected
by the lock of the thunk will have the same values. If this property holds for every execution produced by \RFlock\
and every thunk invoked in each execution, we say that \RFlock\ ensures idempotence.
We prove the following:

\begin{lemma}
\label{lem:idempotent}
\RFlock\ ensures idempotence.
\end{lemma}

We also argue that \RFlock\ does not over-allocate memory or retire the same allocated chunk of memory multiple times.
Roughly speaking, we show that for each chunk of memory that the threads attempt to allocate, only one thread records a pointer to its
allocated chunk in the read log and the rest use that version (quitting from using the versions they allocated).
Similarly, at most one thread retires each allocated chunk of memory.

Consider an execution $\alpha$.
An instance $I$ of \TryLock\ executed in $\alpha$ is successful if it returns TRUE and unsuccessful if it returns FALSE.
(A nested lock succeeds if the outer lock and all locks nested in it succeed.)
Intuitively, $I$ is correct  if the following holds: if $I$ is unsuccessful, none of the updates performed in the thunk(s) it executes
has effect, whereas if it is successful, then all the updates performed in the thunk(s) it executes must have effect.
We say that \RFlock\ is correct if in every exeution it produces, all instances of \TryLock\ that are invoked
in the execution, are correct.
We argue the following:

\begin{lemma}
\label{sec:trylock}
\RFlock\  is a correct agorithm.
\end{lemma}

For studying progress, we make similar assumtions as in~\cite{ben2022lock}.
We first assume that all locks have a partial order
and that the acquisition of nested locks respects this order.
We also assume that the number of locks is bounded and that the time a thread holds a lock is bounded~\cite{ben2022lock}.
We argue that \RFlock\ is lock-free.

\begin{lemma}
In every infinite execution produced by \RFlock,
which contains a finite number of system crashes,
an infinite number of invocations to \TryLock\ completes.
\end{lemma}

\section{Nested Locks in \RFlock}
\label{sec:nesting}

We now briefly describe how \RFlock\ handles nested critical sections.
In Algorithms~\ref{alg:rec-no-nest:mutable}-\ref{alg:rec-no-nest:try-lock}, we highlight in red
the part of the pseudocode which copes with nesting. 

\RFlock\ implements updates in a deferred way
for achieving good performance.
To respect this design decision, nested locking
is implemented so that all updates performed in nested critical sections 
occur at the end of the
outermost critical section and  all nested locks are released together
with the outermost lock, i.e., nesting requires that the original program 
follows two-phase locking. 
\RFlock\ could easily support in-place updates,
and thus it can also support nested locking without making
this assumption. However, this would violate well-known persistence principles~\cite{AB+22,FKK22},
increasing the number of persistence instructions that are needed, which would
result in higher persistence cost.

The above design decision of \RFlock\ for nested locks requires that
when a thread $q$ wants to reserve an acquired nested lock $L'$,
it has to help $L'$'s owner thread $p$
to finish the outermost critical section $L$ that contains $L'$, as well as all critical sections that are nested in $L$.
This requires mechanisms that allow $q$ to figure out that $L'$ is nested and discover $p$ and $L$.
Moreover, to avoid circular helping, $q$ must remember that it helps $p$, so that
it does not restart the helping process (starting from the outermost critical section) each time it acquires one of $L$'s nested locks.
\RFlock\ maintains the lock log and many variables per thread to handle these issues.
For instance, while $q$ executes a thunk, variable $\mathit{wth}$, which is initially $q$,
is updated to $p$ each time $q$ helps a thread $p$ (lines 87-91 of \Run).

To execute nested critical sections, we re-use the logs of the Lock protecting the outermost critical section.
\CreateDescriptor\ has been updated to accommodate this change.
Moreover, we now use an extra log, namely the lock log,
where all locks are recorded. This is also needed for releasing all locks at the end of the outermost critical section.
\RFlock\ maintains also two variables, $\mathit{topD}[q]$ and $\mathit{topL}[q]$, for each thread $q$,
where $q$ records pointers to the descriptor and the lockDescr of the outermost lock of its current critical section
at each point in time. If $q$ does not execute any critical section at $t$, then these pointers are \NULL.

\noindent{\bf Recording Nested-Lock Descriptors in Read-Logs.}
Consider a critical section $\mathit{CS}$ that is not nested
and let $\mathit{CS'}$ be a critical section  that is nested under $\mathit{CS}$.
Then, \RFlock\ ensures  that the read log of $\mathit{CS}$ contains the LockDescr, $\mathit{ld}$, used to acquire
the lock for $\mathit{CS'}$.
This is needed for the following reason. Assume that several threads
are concurrently trying to acquire the lock for $\mathit{CS'}$ while executing $\mathit{CS}$.
They will all invoke \TryLock\ for $\mathit{CS'}$,
but only one of them will execute the \CAM\
of line~\ref{alg:rflock-no-nest:trylock:lock-acq}
successfully.
By the code, this thread will next invoke \ld\ on the acquired Lock (for $\mathit{CS'}$) on line~\ref{alg:rflock-no-nest:trylock:load}.
Thus, the value of the Lock will be logged in the read log for $\mathit{CS}$.
A helper thread, executing $\mathit{CS}$, will retrieve $\mathit{ld}$ from the read log,
and replay the code for $\mathit{CS'}$ in an idempotent way.

\noindent
{\bf Manipulation of $\mathit{log_q}$.}
To support nested locks, before executing a (possibly nested) thunk (line~\ref{alg:primitives-no-nest:run:thunk}),
a thread $q$ halts and records the currently executing critical section (if any) by calling \SaveLogs\ (line~\ref{alg:primitives-no-nest:save-log}).
\SaveLogs\ stores $q$'s log pointers, and their corresponding active positions
to appropriate private variables (lines~\ref{alg:savelogs:1}-\ref{alg:savelogs:2}).
After the thunk of the (nested) critical section is executed, $q$ resumes the execution of the (outer)
critical section (if any) by re-setting 
these variables to their previous values. This is done by calling \RestoreLogs\ (line~\ref{alg:rflock-no-nest:rununlock:restore}).

\section{Related Work}
\label{sec:related}

Designing lock-free concurrent algorithms, especially for
data structures, have been a vibrant research field.
Shortly after defining the concept of lock-free algorithms,
Herlihy~\cite{Herlihy1993methodology} proposed a methodology
for deriving concurrent lock-free implementations of objects,
from their \emph{sequential} implementation.
Since they are derived from implementations that do not expose
concurrency, the resulting concurrent data structures tend to
be overly synchronized.

Barnes~\cite{DBLP:conf/spaa/Barnes93} present the \emph{cooperative}
method for implementing lock-free data structures based on
a sequential implementation of the data structure.
Each implemented operation is associated with an \emph{operation
descriptor}, that tracks the progress of an operation.
When an operation $\mathit{op}_1$ needs to gain control of a memory
object, it tries to install its descriptor in this object;
if this succeeds, in a sense, the operation holds a \emph{soft}
lock on the object.
This lock is considered \emph{soft},
since another operation $\mathit{op}_2$ needing the same object
will help $\mathit{op}_1$ to complete
(using information in its descriptor).

A similar method is \emph{locking without blocking} suggested
by Turek, Shasha and Prakash~\cite{Turek1992locking}.
Their transformation is applied to algorithms that synchronize
solely by using locks, and avoid deadlock.
The base lock-based algorithms expose situations where processes
can proceed concurrently---when they do not compete for the same locks.

Ten years later, Rajwar and Goodman~\cite{RajwarG2002transactional}
proposed \emph{Transactional Lock Removal} as a way to
transparently execute lock-based synchronization in a lock-free manner.
The idea, inspired by transactional memory~\cite{HerlihyM1993}
and speculative lock elision~\cite{RajwarG2001speculative},
is as follows:
(1) speculatively execute the code sequence protected by a lock,
without trying to acquire this lock, and then
(2) use a conflict resolution mechanism to arbitrate between
conflicting code sequences, and commit the successful one.

A large number of follow-up papers have applied similar ideas to
specific data structures, rather than as general transformations.

The idea of soft, lock-free locks has resurfaced recently~\cite{ben2022lock},
in work that elegantly combines a generic interface
with an efficient implementation.
Their algorithmic ideas have been described in Section~\ref{sec:rflock-over}.

The advent of persistent memory has lead to an interest in lock-free
algorithms that support recovery from failure.
In addition to many specific recoverable and lock-free data structures,
several transformations were suggested to make lock-free algorithms
recoverable.

The \emph{tracking} approach~\cite{attiya2021tracking} is perhaps
the most related to the research thread described earlier.
It incorporates recoverability to lock-free algorithms,
as it assumes that algorithms incorporate \emph{info-objects},
similar in functionality to the object descriptors used,
e.g., in~\cite{DBLP:conf/spaa/Barnes93}.
The info-objects are used to track the progress of an operation,
which in turn, is used to recover its state and effects
during recovery.

\emph{Mirror}~\cite{FriedmanPR2021mirror} is a transformation that
automatically adds durability to lock-free data structures.
The main idea underlying Mirror is to have two copies of the data.
Only the write to the first copy is persisted, while reads are
done from the second copy, which is placed in volatile memory.
Mirror demonstrates good performance in read-dominated workloads.

\emph{FliT}~\cite{WeiBDFBP2022flit} is a library that makes any
linearizable data structure durable,
through the use of a persistent programming interface.
It builds on the generic transformation of
Izraelevits et al.~\cite{DBLP:conf/wdag/IzraelevitzMS16},
but uses a sophisticated tagging mechanism to reduce the number of
flush instructions and achieves good performance.
Lock-freedom is preserved, if the original program was lock-free.
(See a formal verification of FliT in~\cite{BodenmullerDDSW2024}.)

The \emph{Capsules} approach~\cite{BBFW2019} transforms a concurrent
program with reads, writes and CASs into a persistent one,
with a constant overhead.
(See also the implementation of persistent operations
in~\cite{AttiyaBH-PODC2018}.)
Like FliT, this approach preserves the lock-freedom properties
of the program, but a blocking implementation will remain so.

\emph{Memento}~\cite{ChoSRK2023memento} is a general programming
framework for detectably recoverable concurrent data structures,
which preserves lock-freedom.
It provides timestamp-based checkpoint and compare-and-swap operations.
Similarly, \emph{Mangosteen}~\cite{EgorovCDOK2024Mangosteen} is
a programming framework that translates a linearizable application
to be a durably linearizable.
It employs binary instrumentation and redo logging,
and it batches memory accesses to reduce the cost of persistence.

None of the above transformations introduces both lock-freedom and durability
starting from a lock-based implementation.

\section{Conclusion}
\label{sec:conclusion}

In this paper, we proposed \RFlock,
the first transformation that can be applied directly on
a lock-based programme to introduce both
lock-freedom and recoverability. \RFlock\ is general enough to be widely applicable
(supporting also nested locks).
The transformation ensures recoverability without jeopardising
the correctness of the lock-based implementation it is applied on.
We have performed a preliminary set of experiment to study the efficiency of our approach in practice.
The results are promising but a complete experimental analysis is yet to be performed.

\bibliography{bibliography/references}

@inproceedings{WH+20,
author = {Cai, Wentao and Wen, Haosen and Beadle, H. Alan and Kjellqvist, Chris and Hedayati, Mohammad and Scott, Michael L.},
title = {Understanding and optimizing persistent memory allocation},
year = {2020},
isbn = {9781450375665},
publisher = {Association for Computing Machinery},
address = {New York, NY, USA},
booktitle = {Proceedings of the 2020 ACM SIGPLAN International Symposium on Memory Management},
pages = {60--73},
numpages = {14},
location = {London, UK},
series = {ISMM 2020}
}

@inproceedings{AttiyaBH-PODC2018,
	author    = {Hagit Attiya and
		Ohad Ben{-}Baruch and
		Danny Hendler},
	title     = {Nesting-Safe Recoverable Linearizability: Modular Constructions for
		Non-Volatile Memory},
	booktitle = {Proceedings of the 2018 {ACM} Symposium on Principles of Distributed
		Computing, {PODC} 2018, Egham, United Kingdom, July 23-27, 2018},
	pages     = {7--16},
	year      = {2018},
	bibsource = {dblp computer science bibliography, https://dblp.org}
}

@article{DBLP:journals/corr/abs-1806-04780,
	author    = {Naama Ben{-}David and
		Guy E. Blelloch and
		Yuanhao Wei},
	title     = {Making Concurrent Algorithms Detectable},
	journal   = {CoRR},
	volume    = {abs/1806.04780},
	year      = {2018},
	url       = {http://arxiv.org/abs/1806.04780},
	bibsource = {dblp computer science bibliography, https://dblp.org}
}

@inproceedings{DBLP:conf/opodis/BerryhillGT15,
	author    = {Ryan Berryhill and
		Wojciech M. Golab and
		Mahesh Tripunitara},
	title     = {Robust Shared Objects for Non-Volatile Main Memory},
	booktitle = {19th International Conference on Principles of Distributed Systems,
		{OPODIS} 2015, December 14-17, 2015, Rennes, France},
	pages     = {20:1--20:17},
	year      = {2015},
	url       = {https://doi.org/10.4230/LIPIcs.OPODIS.2015.20},
	bibsource = {dblp computer science bibliography, https://dblp.org}
}

@article{ChenQ-VLDB2015,
	author    = {Shimin Chen and
		Qin Jin},
	title     = {Persistent B+-Trees in Non-Volatile Main Memory},
	journal   = {{PVLDB}},
	volume    = {8},
	number    = {7},
	pages     = {786--797},
	year      = {2015},
	url       = {http://www.vldb.org/pvldb/vol8/p786-chen.pdf},
	bibsource = {dblp computer science bibliography, https://dblp.org}
}

@inproceedings{CoburnCAGGJW-Asplos2011,
	author    = {Joel Coburn and
		Adrian M. Caulfield and
		Ameen Akel and
		Laura M. Grupp and
		Rajesh K. Gupta and
		Ranjit Jhala and
		Steven Swanson},
	title     = {NV-Heaps: making persistent objects fast and safe with next-generation,
		non-volatile memories},
	booktitle = {Proceedings of the 16th International Conference on Architectural
		Support for Programming Languages and Operating Systems, {ASPLOS}
		2011, Newport Beach, CA, USA, March 5-11, 2011},
	pages     = {105--118},
	year      = {2011},
	url       = {http://doi.acm.org/10.1145/1950365.1950380},
	bibsource = {dblp computer science bibliography, https://dblp.org}
}

@inproceedings{DBLP:conf/ppopp/FriedmanHMP18,
	author    = {Michal Friedman and
		Maurice Herlihy and
		Virendra J. Marathe and
		Erez Petrank},
	title     = {A persistent lock-free queue for non-volatile memory},
	booktitle = {Proceedings of the 23rd {ACM} {SIGPLAN} Symposium on Principles and
		Practice of Parallel Programming, PPoPP 2018, Vienna, Austria, February
		24-28, 2018},
	pages     = {28--40},
	year      = {2018},
	url       = {http://doi.acm.org/10.1145/3178487.3178490},
	bibsource = {dblp computer science bibliography, https://dblp.org}
}

@book{DBLP:books/daglib/0020056,
	author    = {Maurice Herlihy \&
		Nir Shavit},
	title     = {The art of multiprocessor programming},
	publisher = {Morgan Kaufmann},
	year      = {2008},
	isbn      = {978-0-12-370591-4},
	bibsource = {dblp computer science bibliography, https://dblp.org}
}

@inproceedings{DBLP:conf/wdag/IzraelevitzMS16,
	author    = {Joseph Izraelevitz and
		Hammurabi Mendes and
		Michael L. Scott},
	title     = {Linearizability of Persistent Memory Objects Under a Full-System-Crash
		Failure Model},
	booktitle = {Distributed Computing - 30th International Symposium, {DISC} 2016,
		Paris, France, September 27-29, 2016. Proceedings},
	pages     = {313--327},
	year      = {2016},
	url       = {https://doi.org/10.1007/978-3-662-53426-7\_23},
	bibsource = {dblp computer science bibliography, https://dblp.org}
}

@inproceedings{MichaelS-PODC1996,
	author    = {Maged M. Michael and
		Michael L. Scott},
	title     = {Simple, Fast, and Practical Non-Blocking and Blocking Concurrent Queue
		Algorithms},
	booktitle = {Proceedings of the 15th ACM Symposium on Principles
		of Distributed Computing, Philadelphia, Pennsylvania, USA, May 23-26,
		1996},
	pages     = {267--275},
	year      = {1996},
	url       = {http://doi.acm.org/10.1145/248052.248106},
	bibsource = {dblp computer science bibliography, https://dblp.org}
}

@inproceedings{DBLP:conf/spaa/Barnes93,
  author    = {Greg Barnes},
  title     = {A Method for Implementing Lock-Free Shared-Data Structures},
  booktitle = {Proceedings of the 5th Annual {ACM} Symposium on Parallel Algorithms
               and Architectures, {SPAA} '93, Velen, Germany, June 30 - July 2, 1993},
  pages     = {261--270},
  year      = {1993},
  bibsource = {dblp computer science bibliography, https://dblp.org}
}

@inproceedings{Turek1992locking,
  title={Locking without blocking: making lock based concurrent data structure algorithms nonblocking},
  author={Turek, John and Shasha, Dennis and Prakash, Sundeep},
  booktitle={Proceedings of the 11th ACM  Symposium on Principles of database systems},
  pages={212--222},
  year={1992}
}

@inproceedings{BBFW2019,
  title={Delay-free concurrency on faulty persistent memory},
  author={Ben-David, Naama and Blelloch, Guy E and Friedman, Michal and Wei, Yuanhao},
  booktitle={The 31st ACM Symposium on Parallelism in Algorithms and Architectures},
  pages={253--264},
  year={2019}
}

@inproceedings{RamalheteCFC19,
	author    = {Pedro Ramalhete and
	Andreia Correia and
	Pascal Felber and
	Nachshon Cohen},
	title     = {OneFile: {A} Wait-Free Persistent Transactional Memory},
	booktitle = {49th Annual {IEEE/IFIP} Conference on Dependable Systems
	and Networks, {DSN} 2019, Portland, OR, USA, June 24-27, 2019},
	pages     = {151--163},
	publisher = {{IEEE}},
	year      = {2019},
	url       = {https://doi.org/10.1109/DSN.2019.00028},
	bibsource = {dblp computer science bibliography, https://dblp.org}
}

@inproceedings{FB+20,
author = {Friedman, Michal and Ben-David, Naama and Wei, Yuanhao and Blelloch, Guy E. and Petrank, Erez},
title = {NVTraverse: In NVRAM Data Structures, the Destination is More Important than the Journey},
year = {2020},
isbn = {9781450376136},
publisher = {Association for Computing Machinery},
address = {New York, NY, USA},
url = {https://doi.org/10.1145/3385412.3386031},
booktitle = {Proceedings of the 41st ACM  Conference on Programming Language Design and Implementation},
pages = {377–392},
numpages = {16},
location = {London, UK},
series = {PLDI 2020}
}

@inproceedings{FPR21,
author = {Friedman, Michal and Petrank, Erez and Ramalhete, Pedro},
title = {Mirror: Making Lock-Free Data Structures Persistent},
year = {2021},
isbn = {9781450383912},
publisher = {Association for Computing Machinery},
address = {New York, NY, USA},
url = {https://doi.org/10.1145/3453483.3454105},
booktitle = {Proceedings of the 42nd ACM  Conference on Programming Language Design and Implementation},
pages = {1218–1232},
numpages = {15},
location = {Virtual, Canada},
series = {PLDI 2021}
}

@inbook{SP21,
author = {Sela, Gal and Petrank, Erez},
title = {Durable Queues: The Second Amendment},
year = {2021},
isbn = {9781450380706},
publisher = {Association for Computing Machinery},
address = {New York, NY, USA},
url = {https://doi.org/10.1145/3409964.3461791},
booktitle = {Proceedings of the 33rd ACM Symposium on Parallelism in Algorithms and Architectures},
pages = {385–397},
numpages = {13}
}

@inproceedings{FPR19,
  author    = {Panagiota Fatourou and
               Elias Papavasileiou and
               Eric Ruppert},
  title     = {Persistent Non-Blocking Binary Search Trees Supporting Wait-Free Range
               Queries},
  booktitle = {Proc.\ 31st {ACM} on Symposium on Parallelism in Algorithms and Architectures},
  pages     = {275--286},
  year      = {2019}
}

@inproceedings{IMS16,					
author = {Izraelevitz, Joseph and Mendes, Hammurabi and Scott, Michael L.},
title = {Linearizability of Persistent Memory Objects Under a Full-System-Crash Failure Model},
year = {2016},
publisher = {Springer},
booktitle = {Proceedings of the 30th International Symposium of Distributed Computing},
volume	= {LNCS 9888},
pages = {313--327},
location = {Vienna, Austria},
}

@misc{attiya2021tracking,
      title={Tracking in Order to Recover: Detectable Recovery of Lock-Free Data Structures},
      author={Hagit Attiya and Ohad Ben-Baruch and Panagiota Fatourou and Danny Hendler and Eleftherios Kosmas},
      year={2022},
}

@inproceedings{AB+20,
author = {Attiya, Hagit and Ben-Baruch, Ohad and Fatourou, Panagiota and Hendler, Danny and Kosmas, Eleftherios},
title = {Tracking in Order to Recover - Detectable Recovery of Lock-Free Data Structures},
booktitle = {Proceedings of the 32nd ACM Symposium on Parallelism in Algorithms and Architectures (SPAA)},
year = {2020},
isbn = {9781450369350},
publisher = {Association for Computing Machinery},
address = {New York, NY, USA},
pages = {503--505},
numpages = {3},
location = {Virtual Event, USA},
series = {SPAA '20'}
}

@inproceedings{ben2022lock,
  title={Lock-free locks revisited},
  author={Ben-David, Naama and Blelloch, Guy E and Wei, Yuanhao},
  booktitle={Proceedings of the 27th ACM Symposium on Principles and Practice of Parallel Programming},
  pages={278--293},
  year={2022}
}

@article{ZF+19,
author = {Zuriel, Yoav and Friedman, Michal and Sheffi, Gali and Cohen, Nachshon and Petrank, Erez},
title = {Efficient Lock-Free Durable Sets},
year = {2019},
issue_date = {October 2019},
publisher = {Association for Computing Machinery},
address = {New York, NY, USA},
volume = {3},
number = {OOPSLA},
journal = {Proc. ACM Program. Lang.},
month = {oct},
articleno = {128},
numpages = {26}
}

@inproceedings{FKK22,
author = {Fatourou, Panagiota and Kallimanis, Nikolaos D. and Kosmas, Eleftherios},
title = {The Performance Power of Software Combining in Persistence},
year = {2022},
isbn = {9781450392044},
publisher = {Association for Computing Machinery},
address = {New York, NY, USA},
url = {https://doi.org/10.1145/3503221.3508426},
booktitle = {Proceedings of the 27th ACM Symposium on Principles and Practice of Parallel Programming},
pages = {337–352},
numpages = {16},
location = {Seoul, Republic of Korea},
series = {PPoPP '22}
}

@inproceedings{FriedmanPR2021mirror,
  title={Mirror: making lock-free data structures persistent},
  author={Friedman, Michal and Petrank, Erez and Ramalhete, Pedro},
  booktitle={Proceedings of the 42nd ACM  Conference on Programming Language Design and Implementation},
  pages={1218--1232},
  year={2021}
}

@inproceedings{WeiBDFBP2022flit,
  title={{FliT}: a library for simple and efficient persistent algorithms},
  author={Wei, Yuanhao and Ben-David, Naama and Friedman, Michal and Blelloch, Guy E and Petrank, Erez},
  booktitle={Proceedings of the 27th ACM Symposium on Principles and Practice of Parallel Programming},
  pages={309--321},
  year={2022}
}

@article{RajwarG2002transactional,
  title={Transactional lock-free execution of lock-based programs},
  author={Rajwar, Ravi and Goodman, James R},
  journal={ACM SIGOPS Operating Systems Review},
  volume={36},
  number={5},
  pages={5--17},
  year={2002},
  publisher={ACM New York, NY, USA}
}

@article{Herlihy1993methodology,
  title={A methodology for implementing highly concurrent data objects},
  author={Herlihy, Maurice},
  journal={ACM Transactions on Programming Languages and Systems (TOPLAS)},
  volume={15},
  number={5},
  pages={745--770},
  year={1993},
  publisher={ACM New York, NY, USA}
}

@article{HerlihyM1993,
author = {Herlihy, Maurice and Moss, J. Eliot B.},
title = {Transactional Memory: Architectural Support for Lock-Free Data Structures},
year = {1993},
issue_date = {May 1993},
publisher = {Association for Computing Machinery},
address = {New York, NY, USA},
volume = {21},
number = {2},
issn = {0163-5964},
journal = {SIGARCH Comput. Archit. News},
month = {may},
pages = {289–300},
numpages = {12}
}

@inproceedings{RajwarG2001speculative,
  title={Speculative lock elision: Enabling highly concurrent multithreaded execution},
  author={Rajwar, Ravi and Goodman, James R},
  booktitle={Proceedings. 34th ACM/IEEE International Symposium on Microarchitecture. MICRO-34},
  pages={294--305},
  year={2001},
  organization={IEEE}
}

@article{Herlihy1991wait,
  title={Wait-free synchronization},
  author={Herlihy, Maurice},
  journal={ACM Transactions on Programming Languages and Systems (TOPLAS)},
  volume={13},
  number={1},
  pages={124--149},
  year={1991},
  publisher={ACM New York, NY, USA}
}

@inproceedings{DeKruijfSJ2012static,
  title={Static analysis and compiler design for idempotent processing},
  author={De Kruijf, Marc A and Sankaralingam, Karthikeyan and Jha, Somesh},
  booktitle={Proceedings of the 33rd ACM  Conference on Programming Language Design and Implementation},
  pages={475--486},
  year={2012}
}

@article{Ingerman1961thunks,
  title={Thunks: a way of compiling procedure statements with some comments on procedure declarations},
  author={Ingerman, Peter Zilahy},
  journal={Communications of the ACM},
  volume={4},
  number={1},
  pages={55--58},
  year={1961},
  publisher={ACM New York, NY, USA}
}

@InProceedings{BodenmullerDDSW2024,
author={Bodenm{\"u}ller, Stefan
and Derrick, John
and Dongol, Brijesh
and Schellhorn, Gerhard
and Wehrheim, Heike},
title={A Fully Verified Persistency Library},
booktitle={Verification, Model Checking, and Abstract Interpretation (VMCAI)},
year="2024",
pages="26--47",
}

@inproceedings{EgorovCDOK2024Mangosteen,
author = {Sergey Egorov and Gregory Chockler and Brijesh Dongol and Dan O{\textquoteright}Keeffe and Sadegh Keshavarzi},
title = {{Mangosteen}: {Fast} Transparent Durability for Linearizable Applications using {NVM}},
booktitle = {USENIX Annual Technical Conference (ATC)},
year = {2024},
}

@article{ChoSRK2023memento,
  title={{Memento}: {A} framework for detectable recoverability in persistent memory},
  author={Cho, Kyeongmin and Jeon, Seungmin and Raad, Azalea and Kang, Jeehoon},
  journal={Proceedings of the ACM on Programming Languages},
  volume={PLDI},
  pages={292--317},
  year={2023},
}

@inproceedings{AB+22,
author = {Attiya, Hagit and Ben-Baruch, Ohad and Fatourou, Panagiota and Hendler, Danny and Kosmas, Eleftherios},
title = {Detectable recovery of lock-free data structures},
year = {2022},
isbn = {9781450392044},
publisher = {Association for Computing Machinery},
address = {New York, NY, USA},
booktitle = {Proceedings of the 27th ACM Symposium on Principles and Practice of Parallel Programming},
pages = {262–277},
numpages = {16},
location = {Seoul, Republic of Korea},
series = {PPoPP '22}
}

\end{document}